\documentclass[prb,preprint,showpacs,showkeys]{revtex4}

\usepackage{amssymb}

\def\beq{\begin{equation}}
\def\eeq{\end{equation}}

\def\rmd{{\rm d}}

\catcode`@=11  
\def\meqalign#1{\null\,\vcenter{\openup\jot\m@th
  \ialign{\strut\hfil$\displaystyle{##}$&&$\displaystyle{{}##}$\hfil
     \crcr#1\crcr}}\,}
\catcode`@=12  

\def\pmb#1{\setbox0=\hbox{$#1$}%
  \kern-.025em\copy0\kern-\wd0
  \kern.05em\copy0\kern-\wd0
  \kern-.025em\raise.0433em\box0}

\begin{document}

\title{Massless field perturbations of the spinning C metric}

\author{D. Bini}\thanks{Electronic mail: binid@icra.it}
\affiliation{Istituto per le Applicazioni del Calcolo ``M. Picone,'' CNR I-00161 Rome, Italy}
\affiliation{ICRA, University of Rome ``La Sapienza,'' I-00185 Rome, Italy}
\affiliation{INFN - Sezione di Firenze, Polo Scientifico, Via Sansone 1, I-50019, Sesto Fiorentino (FI), Italy}

\author{C. Cherubini}\thanks{Electronic mail: cherubini@icra.it}
\affiliation{Facolt\`a di Ingegneria, Universit\`a Campus Biomedico, Via Alvaro del Portillo 21,  I-00128 Roma, Italy}
\affiliation{ICRA, University of Rome ``La Sapienza,'' I-00185 Rome, Italy}

\author{A. Geralico}\thanks{Electronic mail: geralico@icra.it}
\affiliation{Physics Department and ICRA, University of Rome ``La Sapienza,'' I-00185 Rome, Italy}

\begin{abstract}
A single master equation is given describing spin $s\le2$ test fields
that are 
gauge- and tetrad-invariant perturbations of the {\it spinning C metric} spacetime representing a source with mass $M$,  uniformly rotating with angular momentum per unit mass $a$ and uniformly accelerated with acceleration $A$.
This equation can be separated into its radial and angular parts. The behavior of the radial functions  near the horizons is studied and used to examine the influence of
$A$ on the phenomenon of superradiance, while the angular equation leads to modified spin-weighted spheroidal harmonic solutions generalizing those of the Kerr spacetime.
Finally the coupling between the spin of the perturbing field and the acceleration parameter $A$ is discussed.
\end{abstract}


\maketitle

\section{Introduction}

The spinning C metric is a boost-rotation-symmetric stationary spacetime of Petrov type D belonging to the Weyl class of solutions of the Einstein equations \cite{ES}. It  can be interpreted as the field of two rotating black holes which are uniformly accelerated in opposite directions under the action of conical singularities.
Using the Boyer-Lindquist-type coordinates $(t,r,\theta,\phi)$ introduced by Podolsky and Griffiths \cite{podgriff1,podgriff2}, the corresponding line element is given by
\begin{eqnarray}
\label{SCmmetric}
\rmd s^2&=& \frac{1}{\Omega^2}\bigg\{\frac1{\Sigma}(Q-a^2P\sin^2\theta)\, \rmd t^2 - \frac{2a\sin^2\theta}{\Sigma} [Q-P(r^2+a^2)]\, \rmd t \,\rmd \phi\nonumber\\
&& - \frac{\sin^2\theta}{\Sigma} [P(r^2+a^2)^2-a^2Q\sin^2\theta]\,\rmd \phi^2 -  \frac{\Sigma}{Q}\,\rmd r^2-\frac{\Sigma}{P}\,\rmd \theta^2\bigg\}\ ,
\end{eqnarray}
where the functions $\Omega$, $\Sigma$, $P$ and $Q$ are defined by
\begin{eqnarray}
\Omega&=&1-Ar\cos\theta\ , \qquad 
\Sigma=r^2+a^2\cos^2\theta\ , \nonumber\\ 
P&=&1-2A{\mathcal M}\cos\theta+a^2A^2\cos^2\theta\ ,  \nonumber\\
Q&=&\Delta(1-A^2r^2)\ , \qquad 
\Delta=r^2-2{\mathcal M}r+a^2=(r-r_-)(r-r_+)\ .
\end{eqnarray}
We list in Appendix A the most important properties of this metric.
Units are chosen such that $G=1=c$, so that the parameters $({\mathcal M},a)$ have the dimension of length whereas $A$ of the inverse of length and  
$$
\bar a=a/{\mathcal M} , \qquad \bar A=A {\mathcal M}
$$
are nondimensional quantities.

In the limit of vanishing rotation parameter $a=0$ the metric (\ref{SCmmetric}) reduces to the usual non-rotating vacuum C metric, while for $A=0$ we recover the familiar Kerr metric for a rotating source; finally, for either $a=0=A$ or $a=0={\mathcal M}$ we get the Schwarzschild and Rindler solutions respectively.

For a further use it is convenient to introduce the notation  
\begin{eqnarray}
\sigma_+&\equiv& (1+a^2A^2)+2{\mathcal M}A=P(\pi), \nonumber \\
\sigma_0&\equiv& (1+a^2A^2)=[P(\pi)+P(0)]/2, \nonumber \\
\sigma_-&\equiv& (1+a^2A^2)-2{\mathcal M}A=P(0).
\end{eqnarray} 

We limit our attention to the spherical corona $r_+<r<r_A=1/A$ in which $Q(r)>0$, also implying $P(\theta)>0$ for all $\theta \in [0,\pi]$.
Finally, we require $\phi\in[0,\phi_0)$, with either $\phi_0=2\pi/\sigma_+$ (equivalent to the removal of the conical singularity at $\theta=\pi$) or $\phi_0=2\pi/\sigma_-$ (equivalent to the removal of the conical singularity at $\theta=0$) as explained in Appendix A. In this way there is no need to specify if the removed singularity corresponds to $\theta=0$ or $\theta=\pi$.

We study here massless perturbations to the spinning C metric due to fields of any spin following the   
approach of Teukolsky  \cite{Teukolsky,barpre} grounded, in turn,  in the context of the Newman-Penrose formalism \cite{NP,chandra}.
For a Kerr gravitational background Teukolsky found a separable master equation whose eigenfunction solutions essentially solve the problem of the massless perturbations of any spin for the Kerr black hole in terms of gauge- and tetrad-invariant quantities. After that work many authors have discussed generalizations to other interesting spacetimes including the vacuum C metric in standard coordinates \cite{Prestidge}.

We introduce in Section II a master equation for the spinning C metric spacetime whose symmetries allow  
the separation of the equation into radial and angular parts, generalizing some previous results valid for the Kerr spacetimes and the non spinning C metric, and use it to study the question of superradiant scattering modes as well as the various allowed coupling terms between the spin of the perturbing field and the background parameters: mass, rotation and acceleration.
Closed form solutions to the radial equation are extensively discussed in Section IV (algebraically special perturbations) as well as in Appendix B (general Liouvillian solutions).

We finally explicitly give in the Appendix C the equations for null geodesics. As in the non-rotating case of the C metric only geodesics of this kind can be separated, because of the existence of a conformal Killing tensor, which is derived too.

\section{Teukolsky Master Equation (TME)}

Consider the spinning C metric in the form  (\ref{SCmmetric}). A Kinnersley-like null frame \cite{kinnersley} 
\begin{eqnarray}
\label{frame}
l&=& \frac{\Omega^2}{Q} [(r^2+a^2) \partial_t+ \frac{Q}{\Omega}\partial_r +a\partial_\phi]\ , \nonumber \\
n&=& \frac{\Omega}{2Q} [(r^2+a^2) \partial_t- \frac{Q}{\Omega}\partial_r +a\partial_\phi] \ ,  \nonumber \\
m&=& \frac{(r-ia\cos \theta)}{\sqrt{2}\Sigma \sqrt{P}} [i\Omega a \sin \theta  \partial_t+P\partial_\theta+i \frac{\Omega}{\sin \theta}\partial_\phi ]\ ,
\end{eqnarray}
can be introduced to define Newman-Penrose (NP) quantities.
The only nonvanishing Weyl scalar is 
\beq
\label{psi2}
\psi_2=-\frac{(1+iaA){\mathcal M}\Omega^3}{(r-ia\cos\theta)^3}
\eeq
and the only nonvanishing spin coefficients are
\begin{eqnarray}
\label{spincoeff}
\rho&=&\frac{i\Omega(i+aA \cos^2 \theta)}{(r-ia\cos\theta)}\ ,\quad \alpha=\pi-\beta^* +\frac{\sqrt{2P}Ar\sin\theta}{(r-ia\cos\theta)}\ , 
\nonumber\\
\mu&=&\frac{Q}{2\Omega^2 \Sigma}\rho \ ,\quad  \pi=-\frac{\sqrt{P} (r^2A-ia)\sin \theta}{\sqrt{2} (r-ia\cos\theta)^2}\ ,
\nonumber\\
\tau&=& \frac{\sqrt{P}(r^2A-ia)\sin \theta}{\sqrt{2}\Sigma}\ ,\quad \gamma=\mu-\frac{Q_{,r}\Omega -4Q\Omega_{,r}}{4\Omega\Sigma} 
\nonumber\\
 \beta&=& -\frac{\sqrt{P}}{2\sqrt{2}}\cot\theta \left(\frac{\rho^*}{\Omega}+A\cos \theta \right)+
\frac{\Omega}{2\sqrt{2}}\frac{(\sqrt{P})_{,\theta}}{(r+ia\cos\theta)}\ .
\end{eqnarray}

A master equation for the gauge- and tetrad-invariant first-order massless perturbations of any spin in this background can be given starting from the following Newman-Penrose relations for any vacuum type D geometry (here considered with no backreaction) \cite{Teukolsky} 
\begin{eqnarray}
\label{mia1}
&&
\{[D-\rho^{*}+\epsilon^*+\epsilon-2s(\rho +\epsilon)](\Delta+\mu-2 s \gamma)
\\ 
&&
  -[\delta+\pi^{*}-\alpha^{*}+\beta
-2 s(\tau+\beta)] \,(\delta^{*}+\pi-2 s\alpha)
  -2(s-1)(s-1/2)\psi_{2}\}\Psi=0 
\nonumber
\end{eqnarray}
for spin weights $s=1/2,1,2$  and
\begin{eqnarray}\label{mia2}
&&
\{[\Delta-\gamma^{*}+\mu^{*}-\gamma-2 s (\gamma+\mu)](D-\rho-2s\epsilon)
\\ 
&&
-[\delta^{*}-\tau^{*}+\beta^{*}-\alpha
-2 s (\alpha+\pi)](\delta-\tau-2 s \beta)
-2(s+1)(s+1/2)\psi_{2}\}\Psi=0
\nonumber
\end{eqnarray}
for $s=-1/2,-1,-2$. The case $s=\pm 3/2$ can be derived instead  by following the work of 
G\"uven \cite{guven}, which is expressed in the alternative Geroch-Held-Penrose formalism \cite{GHP}. Finally the case $s=0$ is given by
\begin{eqnarray}
\meqalign{
&[D\Delta+\Delta D-\delta^* \delta-\delta\delta^*
+(-\gamma-\gamma^*+\mu+\mu^*)D+(\epsilon+\epsilon^*-\rho^*-\rho)\Delta \cr
&+(-\beta^*-\pi+\alpha+\tau^*)\delta+(-\pi^*+\tau-\beta+\alpha^*)\delta^*]\Psi=0\ .
\cr}
\end{eqnarray}
Note that only in these NP equations has the standard notation for the directional derivatives
$D=l^\mu \partial_\mu$, $\Delta=n^\mu \partial_\mu$ and  $\delta=m^\mu \partial_\mu$ been used and the second of these should not be confused  with the equally standard notation for the metric quantity $\Delta= r^2-2{\mathcal M}r+a^2$ used everywhere else in this article.

As in the case of the Kerr spacetime \cite{bcjr} and the Taub-NUT spacetime \cite{TeukTN} all these equations for distinct spin weights can be cast into a single compact form in the 
spinning C metric spacetime as well, by introducing a ``connection vector" with components
\begin{eqnarray}
&\Gamma^t =& \frac{\Omega^2}{\Sigma}\left\{\frac{1}{Q}\left[{\mathcal M}(A^2r^4+a^2)+r\sigma_0(\Delta -{\mathcal M}r)\right]\right.\nonumber \\
&&\left.+i\frac{a}{P} 
\left[\sigma_0\cos\theta -A{\mathcal M}(1+\cos^2\theta)\right] \right\}\ ,
\nonumber\\
&\Gamma^r =& -\frac{\Omega}{\Sigma}\left( \frac12 \Omega Q_{,r}+2A\cos \theta Q\right)\ ,
\nonumber\\
&\Gamma^\theta =& \frac{2A\Omega  P r\sin \theta}{\Sigma} \ , 
\nonumber\\
&\Gamma^\phi =&-\frac{\Omega^2}{\Sigma}\left[\frac{aQ_{,r}}{2Q}+i\left( \frac{\cos\theta (2P-1)+A{\mathcal M}(\cos^2\theta -A^2a^2 \cos\theta +1)}{P\sin^2 \theta}\right)
\right]\ .
\label{eq:SPINNOL}
\end{eqnarray}
The resulting master equation has the form
\begin{equation}
[(\nabla^\mu+s\Gamma^\mu)(\nabla_\mu+s\Gamma_\mu)-4s^2\psi_2]\Psi=0\ ,\qquad 
{\textstyle s=0,\pm\frac12,\pm1,\pm\frac32,\pm2}\ , 
\label{eq:bellak}
\end{equation}
where $\psi_2$ is the spinning C metric background Weyl scalar given by (\ref{psi2}).
This master equation characterizes the common behavior of all these massless fields in this background differing only in the value of the spin-weight parameter $s$. In fact, the first term on its left-hand side represents (formally) a wave (d'Alembert) operator, corrected by taking into account the spin-weight of the perturbing field, and the second term is a (Weyl) curvature term also linked to the spin-weight value. 
Table \ref{TAB} shows the various Newman-Penrose quantities for which the master equation holds following the standard notation \cite{Teukolsky}, where in the spin-2 case $\psi_0$ and $\psi_4$ refer to the perturbed Weyl scalars.

\begin{table}
\begin{center}
\begin{tabular}{|c||c|c|c|c|c|c|c|c|c|}
\hline
\rule{0pt}{2ex}$s$ & 0 & 1/2 & -1/2 & 1 & -1 & 3/2 & -3/2 & 2 & -2 \\
\hline 
\rule{0pt}{2ex}$\Psi$ & $\Phi$ & $\chi_0$ & $\rho^{-1}\chi_1$ & $\phi_0 $ & $\rho^{-2}\phi_2$ &
$\Omega_0$ & $\rho^{-3}\Omega_3$ & $\psi_0$ & $\rho^{-4}\psi_4$ \\
\hline
\end{tabular}
\end{center}
\caption{The spin-weight $s$ and the physical field component $\Psi$ for the master equation.}
\label{TAB}
\end{table}

\section{Separation of the master equation}

Remarkably the master equation (\ref{eq:bellak}) admits separable solutions of the form
\beq
\psi(t,r,\theta,\phi)=\Omega^{(1+2s)}e^{-i\omega t}e^{i m \phi} R(r)S(\theta)\ ,
\eeq
where $\omega>0$ is the wave frequency and $m$ is the azimuthal separation constant. 
Note that as stated above the conical singularity along the symmetry axis is removed by insisting periodicity in the
azimuthal coordinate such that $\Delta\phi=\phi_0$. 
Since $\phi$ is in this way a periodic coordinate, $m$ must be of the form $m=(2\pi/\phi_0)m_0=m_0\sigma_\pm$, with $m_0$ a positive integer, without loss of generality.

The radial equation is then
\beq
\label{radeq}
Q^{-s} \frac{\rmd}{\rmd r}\left(Q^{s+1}\frac{\rmd R(r)}{\rmd r} \right)
+V_{\rm (rad)}(r)R(r)=0\ ,
\eeq
with
\begin{eqnarray}
\label{radiale1}
V_{\rm (rad)}(r) &=&-2rA^2(r-{\mathcal M})(1+s)(1+2s)+\frac{\kappa(r)^2}{Q} \nonumber \\
&& -2is \left[
-\frac{amQ_{,r}}{2Q}+\frac{\omega {\mathcal M}(r^2-a^2)}{\Delta}-\frac{\omega r \sigma_0}{1-A^2r^2}
\right]+2K
\end{eqnarray}
where $K$ is the separation constant and the quantity
\beq
\kappa(r)=(r^2+a^2)\omega-am
\eeq 
has been introduced following Teukolsky.
Clearly the solution $R(r)$ of this equation depends on the value of the spin weight $s$, so when convenient this dependence will be made explicit using the notation $R(r)\equiv R_s(r)$.
An analytic solution of this equation is expected only for very special values of $K$ (termed \lq\lq eigenvalues"). 
Algebraically special perturbations are considered in Section IV.
The existence of other closed form (or Liouvillian) solutions to this equation obtained by constraining black hole parameters, spacetime acceleration as well as separation constants is discussed in Appendix B.

Equation (\ref{radeq}) will be studied on the interval $r\in (r_+ , r_A )$, where the metric and the chosen tetrad (\ref{frame}) are well behaved, closely following the usual treatment of black hole perturbations that motivates the present investigation. 

By introducing the scaling
\beq
R(r)=(r^2+a^2)^{-\frac 12}Q^{-\frac s2}H(r)\equiv \mathcal{Q}_s^{-1}\, H(r)
\eeq
and the ``tortoise" coordinate transformation $r\to r_*$, where
\begin{eqnarray}
\frac{\rmd r}{\rmd r_*}&=&\frac{Q}{r^2+a^2}\ ,\nonumber \\
r_*&=& \sigma_0\left[\frac{1}{A\sigma_+}\ln \sqrt{1+Ar} -
\frac{1}{A\sigma_-}\ln \sqrt{1-Ar}+\frac{{\mathcal M}}{\sigma_+\sigma_-}\ln \Delta 
 \right]\nonumber \\
&& +\frac{2{\mathcal M}^2(1-a^2A^2)}{\sigma_+\sigma_- \sqrt{{\mathcal M}^2-a^2}}\ln \sqrt{\frac{r-r_+}{r-r_-}}+{\rm const}.\ ,
\end{eqnarray}
the radial equation can be transformed into the one-dimensional Schr\"odinger-like equation
\beq
\label{newradi}
\frac{\rmd ^2 }{\rmd r_*^2}H(r)+\tilde V H(r)=0\ ,
\eeq
with the potential
\begin{eqnarray}
\tilde V&=& \left[\frac{\kappa(r)}{r^2+a^2}-iG\right]^2-\frac{\rmd G}{\rmd r_*}
\nonumber \\
&&-\frac{2Q}{(r^2+a^2)^2}\left[rA^2 (r-{\mathcal M})(1+s)(1+2s)-K-2i\omega rs-\frac{ir\kappa(r) }{(r^2+a^2)}\right]\ ,
\end{eqnarray}
where
\beq
G=\frac{s[(r-{\mathcal M})(1-r^2A^2)-rA^2\Delta]}{(r^2+a^2)}+\frac{rQ}{(r^2+a^2)^2}
=\frac{\rmd}{\rmd r_*} \ln \mathcal{Q}_s
\eeq
has been introduced in analogy with Teukolsky's treatment of the perturbations of the exterior Kerr spacetime.

The asymptotic form of the radial equation as $r\to r_A \, (r_{*}\to\infty)$ is
\beq
\frac{\rmd ^2 }{\rmd r_*^2}H(r)+\left(\kappa_A-i\beta_A \right )^2 H(r)=0\ ,
\eeq
where 
\beq
\beta_A=-\frac{As\sigma_-}{\sigma_0}\ , \qquad\kappa_A=\omega-m\omega_A\ , \qquad \omega_A=\frac{aA^2}{\sigma_0}\ .
\eeq
We notice that for small values of $A$ ($r_A \to \infty $) we approach the Teukolsky result:
\beq
\left(\kappa_A-i\beta_A \right )^2 \to \omega^2+ \frac{2is\omega}{r_A} +O(A^2)\ .
\eeq

On the other hand  close to the horizon $r\to r_+\, (r_{*}\to-\infty)$, the  asymptotic form of the radial equation becomes
\beq
\frac{\rmd ^2 }{\rmd r_*^2}H(r)+\left( \kappa_+ -i \beta_{+} \right )^2 H(r)=0\ ,
\eeq
where 
\beq\label{omegahor} 
\beta_{+}=\frac{s(r_+-{\mathcal M})(1-r_+^2A^2)}{2{\mathcal M} r_+}\ , \quad \kappa_+=\omega -m\omega_+\ , \quad \omega_+=\frac{a}{r_+^2+a^2}=\frac{a}{2{\mathcal M} r_+}\ ,
\eeq 
with  $\omega_+$ the ``effective angular velocity" of the horizon. 
Following Teukolsky, if $\frac{m\omega_+}{\omega}>1$ energy flows out from the hole, i.e. one has superradiant scattering. 
Superradiance (see Ref. \cite{JAC} and references therein for an exaustive review) is a physical effect typically related  to rotation of totally absorbing objects. 
More specifically, Zel'dovich \cite{zeldovich} in 1971 noticed that a cylinder made of absorbing material and rotating about its axis with frequency $\omega_+$ can amplify modes of scalar or electromagnetic radiation scattering on
it which satisfy the condition  $\omega-m\omega_+ < 0$, where $\omega$ is the  frequency of waves and $m$ the usual azimuthal quantum number. 
In particular, in a curved spacetimes or in the effective curved geometries typical of acoustic black holes\cite{chersucci1}, the existence of an ergosphere allows one to extract rotational energy from the central engine, i.e. a field theory version of Penrose's process for point-like particles.
In presence of costant acceleration only (i.e. no rotation), superradiant effects are not expected. 
On the contrary, in presence of rotation but not of constant acceleration, it is well known from Teukolsky's work that superradiant modes depend from the quantity $\omega_+$ previously computed. 
It is natural to question now what happens in the more general case, which is the goal of our work:  
in case of both acceleration $A$ and rotation $a$, related via the nonlinearities in Einstein's theory, the quantity  $\omega_+$ remains the same of Kerr solution, i.e. there is no coupling of acceleration and rotation in superradiant modes. This is a novel and a priori unxepected result of our analysis, and can have relevant implications for the understanding  both of complicated dynamics of moving rotating black holes in numerical relativity, as well as of experiments in ordinary fluids in the context of induced geometries.

To complete our analysis of the perturbations on the spinning C metric spacetime, we need to discuss the angular equation
\beq
\label{angolare}
\frac{1}{\sin\theta}\frac{\rmd }{\rmd \theta} 
  \left(\sin\theta \frac{\rmd Y(\theta)}{\rmd \theta}\right)
  + V_{\rm (ang)}(\theta)Y(\theta)=0\ ,
\eeq
where
\begin{eqnarray}
\label{angolare1}
V_{\rm (ang)}(\theta)
&=&-\frac{2K-s(1-A^2a^2)}{P}+\frac{1}{P^2}\bigg\{
\frac{-\sigma_0^2s^2\cos^2\theta+2\sigma_0sw\cos\theta-w^2\cos^2\theta}{\sin^2\theta}\nonumber\\
&&+z^2\cos^2\theta-2s\sigma_0z\cos\theta-(z+w-4sA{\mathcal M})^2\nonumber\\
&&+A^2[({\mathcal M}^2-a^2)\sin^2\theta+4sa^2\cos\theta(2sA{\mathcal M}-w)]
\bigg\}\ ,
\end{eqnarray}
or equivalently:
\begin{eqnarray}
\label{angolare2}
V_{\rm (ang)}(\theta)
&=&\frac{1-2K+s(2-\sigma_0)}{P}+\frac{1}{P^2}\bigg\{
-\frac{(w\cos\theta-\sigma_0 s)^2}{\sin^2\theta}\nonumber\\
&&-(z+w-4s{\bar A})^2+(z\cos\theta-s\sigma_0)^2-({\bar A}\cos \theta -1)^2\nonumber\\
&&+1-\sigma_0+{\bar A}^2+4s(\sigma_0-1)\cos\theta(2s{\bar A}-w)
\bigg\}\ ,
\end{eqnarray}
where $Y(\theta)=\sqrt{P}S(\theta)$, $z=a\omega+s{\bar A}$ and $w=-m+2s{\bar A}$.
This equation generalizes the spin-weighted spheroidal harmonics of Teukolsky \cite{Teukolsky,FC,novello}.

In the limit of small values of the rotation parameter $a$ as well as acceleration parameter $A$, also neglecting terms of the order $aA$, the radial and angular potentials (\ref{radiale1}) and (\ref{angolare1}) reduce to
\begin{eqnarray}
V_{\rm (rad)}(r) &\simeq&\frac{\omega^2 r^3}{r-2{\mathcal M}} +2is\omega r 
\frac{ r-3{\mathcal M}}{r-2{\mathcal M}}+2K -\frac{2ma}{r(r-2{\mathcal M})}[\omega r^2-is(r-{\mathcal M})]\ , \nonumber\\
V_{\rm (ang)}(\theta)&\simeq&\left[s-2K-\frac{(s\cos\theta+m)^2}{\sin^2\theta}\right](1+2{\bar A}\cos\theta)-2{\bar A}\left[sm-\frac{(s^2-m^2)\cos\theta}{\sin^2\theta}\right]\nonumber\\
&&+2\omega a (m-s\cos\theta)\ ,
\end{eqnarray}
respectively.
Note that the corrections due to $A$ to the radial potential  are $O(A^2)$ while those to the angular potential are $O(A)$.

We will discuss now the limiting cases of Kerr, Schwarzschild, C metric and Rindler spacetimes.

\subsection{Limiting cases}

\subsubsection{Kerr: $A=0$}

The radial equation (\ref{radeq}) reduces to
\beq
\label{radeqkerr}
\Delta^{-s} \frac{\rmd}{\rmd r}\left(\Delta^{s+1}\frac{\rmd R(r)}{\rmd r} \right)
+V_{\rm (rad)}^K(r)R(r)=0\ ,
\eeq
with
\begin{eqnarray}
V_{\rm (rad)}^K(r) &=&\frac{[-am +\omega(r^2+a^2)]^2}{\Delta} \nonumber \\
&& -2is \left[
-\frac{am\Delta_{,r}}{2\Delta}+\frac{\omega {\mathcal M}(r^2-a^2)}{\Delta}-\omega r\right]+2K
\end{eqnarray}
where $K$ is the separation constant. 

The angular equation (\ref{angolare}) becomes
\beq
\frac{1}{\sin\theta}\frac{\rmd }{\rmd \theta} 
  \left(\sin\theta \frac{\rmd Y(\theta)}{\rmd \theta}\right)
  + V_{\rm (ang)}^K(\theta)Y(\theta)=0\ ,
\eeq
where
\begin{eqnarray}
V_{\rm (ang)}^K(\theta)
&=&s-2K-
\frac{(s\cos\theta+m)^2}{\sin^2\theta}\nonumber\\
&&-a^2\omega^2\sin^2\theta-2sa\omega\cos\theta+2a\omega m
\ , \\
&=&s-s^2+2a\omega m-a^2\omega^2-2K-
\frac{(s\cos\theta+m)^2}{\sin^2\theta}\nonumber\\
&&+(a\omega\cos\theta-s)^2
\ ,
\end{eqnarray}
where $Y(\theta)=S(\theta)$.

Following Teukolsky the separation constant in this case should be set as $K=[\Omega+s(s+1)-a\omega(a\omega-2m)]/2$, where $\Omega=-L(L+1)$.

\subsubsection{Schwarzschild: $A=0$, $a=0$}

The radial equation (\ref{radeq}) reduces to
\beq
\label{radeqschw}
(r^2-2{\mathcal M}r)^{-s} \frac{\rmd}{\rmd r}\left((r^2-2{\mathcal M}r)^{s+1}\frac{\rmd R(r)}{\rmd r} \right)
+V_{\rm (rad)}^S(r)R(r)=0\ ,
\eeq
with
\begin{eqnarray}
V_{\rm (rad)}^S(r) &=&\frac{\omega^2 r^3}{r-2{\mathcal M}} +2is\omega r 
\frac{ r-3{\mathcal M}}{r-2{\mathcal M}}+2K
\end{eqnarray}
where $K$ is the separation constant. 

The angular equation (\ref{angolare}) becomes
\beq
\frac{1}{\sin\theta}\frac{\rmd }{\rmd \theta} 
  \left(\sin\theta \frac{\rmd Y(\theta)}{\rmd \theta}\right)
  + V_{\rm (ang)}^S(\theta)Y(\theta)=0\ ,
\eeq
where
\begin{eqnarray}
V_{\rm (ang)}^S(\theta)
&=&s-2K-
\frac{s^2\cos^2\theta+2sm\cos\theta+m^2}{\sin^2\theta}\ ,
\end{eqnarray}
where $Y(\theta)=S(\theta)$.

The separation constant in this case should be set as $K=[\Omega+s(s+1)]/2$, where $\Omega=-L(L+1)$.

\subsubsection{C metric: $a=0$}

The radial equation (\ref{radeq}) reduces to
\begin{eqnarray}
\label{radeqcmet}
&&[(1-A^2r^2)(r^2-2{\mathcal M}r)]^{-s} \frac{\rmd}{\rmd r}\left([(1-A^2r^2)(r^2-2{\mathcal M}r)]^{s+1}\frac{\rmd R(r)}{\rmd r} \right)\nonumber\\
&&+V_{\rm (rad)}^C(r)R(r)=0\ ,
\end{eqnarray}
with
\begin{eqnarray}
V_{\rm (rad)}^C(r) &=&-2rA^2(r-{\mathcal M})(1+s)(1+2s)+\frac{\omega^2r^4}{1-A^2r^2} \nonumber \\
&& -2is\omega r \left(
\frac{{\mathcal M}}{r-2{\mathcal M}}-\frac{1}{1-A^2r^2}
\right)+2K
\end{eqnarray}
where $K$ is the separation constant. 

The angular equation (\ref{angolare}) becomes
\beq
\frac{1}{\sin\theta}\frac{\rmd }{\rmd \theta} 
  \left(\sin\theta \frac{\rmd Y(\theta)}{\rmd \theta}\right)
  + V_{\rm (ang)}^C(\theta)Y(\theta)=0\ ,
\eeq
where
\begin{eqnarray}
V_{\rm (ang)}^C(\theta)
&=&-\frac{2K-s}{P_C}+\frac{1}{P_C^2}\bigg\{
\frac{-s^2\cos^2\theta+2sw\cos\theta-w^2\cos^2\theta}{\sin^2\theta}\nonumber\\
&&+z^2\cos^2\theta-2sz\cos\theta-(m+s{\bar A})^2+{\bar A}^2\sin^2\theta
\bigg\}\ ,
\end{eqnarray}
where $Y(\theta)=\sqrt{P_C}S(\theta)$, $P_C=1-2{\bar A}\cos\theta$, $z=s{\bar A}$ and $w=-m+2s{\bar A}$.

\subsubsection{Rindler: $M=0=a$}

The radial equation (\ref{radeq}) reduces to
\beq
\label{radeqrind}
(1-A^2r^2)^{-s} \frac{\rmd}{\rmd r}\left((1-A^2r^2)^{s+1}\frac{\rmd R(r)}{\rmd r} \right)
+V_{\rm (rad)}^R(r)R(r)=0\ ,
\eeq
with
\begin{eqnarray}
V_{\rm (rad)}^R(r) &=&-2r^2A^2(1+s)(1+2s)+\frac{\omega r(\omega r^3+2is)}{1-A^2r^2}+2K
\end{eqnarray}
where $K$ is the separation constant. 

The angular equation (\ref{angolare}) becomes
\beq
\frac{1}{\sin\theta}\frac{\rmd }{\rmd \theta} 
  \left(\sin\theta \frac{\rmd Y(\theta)}{\rmd \theta}\right)
  + V_{\rm (ang)}^R(\theta)Y(\theta)=0\ ,
\eeq
where
\begin{eqnarray}
V_{\rm (ang)}^R(\theta)
&=&s-2K-\frac{s^2\cos^2\theta+2sm\cos\theta+m^2}{\sin^2\theta}\ ,
\end{eqnarray}
where $Y(\theta)=S(\theta)$.

\section{Algebraically special perturbations}

Consider the radial equation (\ref{radeq}). In analogy with the Schwarzschild black hole case one can find analytic solutions of this equation for $s=+2$ given by
\beq
R(r)=\frac{c_3r^3+c_2r^2+c_1r+c_0}{Q^2}e^{-i\omega \tilde r}, \qquad \tilde r= \int \frac{r^2+a^2-\frac{am}{\omega}}{Q} \, \rmd r\ .
\eeq
Substituting into Eq. (\ref{radeq}) one obtains a 7$^{th}$ degree polynomial equation in $r$ which must be identically zero, i.e. the coefficient of any power of $r$ should vanish leading to a homogeneous linear system of $8$ equations in the $4$ variables
$c_0, \ldots c_3$ (consequence of the fact that the coefficient of any power of $r$ is a linear function of  $c_0, \ldots c_3$).
This system is compatible assuming the following constraint (generalizing the Starobinsky constraint):
\beq
P_2(x)+P_4(y)+P_2^q(y)+C=0\ ,
\eeq
where $x={\mathcal M}\omega$, $y=a\omega$, $q={\bar a}{\bar A}$ and
\begin{eqnarray}
P_2(x)&=&9x^2 \nonumber \\
P_4(y)&=&9y^2(y-m)^2-2y(K-2)[(5K-7)y -(5K-13)m] \nonumber \\
P_2^q(y)&=& 9(y-m)^2q^2\bar A^2-(28q^2-74+34K)q^2y^2+2q^2my (2q^2+11K-37).
\end{eqnarray}
We have 
\begin{eqnarray}
C&=& [(K-2)(K-3)-3{\bar A}^2+q^2(5K-9+6q^2)]^2\nonumber\\
&&+3m^2q^2(8q^2+4K-8)\nonumber \\
&\equiv & W^2 +12m^2q^2[K-2(1-q^2)] \ .
\end{eqnarray}

The solution is implicitly given by
\begin{eqnarray}
\frac{c_0}{{\mathcal M}^4}&=& \frac{c_1}{ {\mathcal M}^3} \frac{[i\bar a(y-m)-1]}{K-2(1-q^2)}-\frac{c_2}{ {\mathcal M}^2} \frac{\bar a^2}{K-2(1-q^2)}\nonumber \\
\frac{c_3}{{\mathcal M}}&=& \frac{c_1}{{\mathcal M}^3} \frac{\bar A^2}{K-2(1-q^2)}+\frac{c_2}{{\mathcal M}^2} \frac{\bar A^2-x}{K-2(1-q^2)}
\end{eqnarray}
with $c_1$ arbitrary and $c_2$ given by
\begin{eqnarray}
\frac{c_2}{{\mathcal M}^2}&=&2i\frac{c_1}{ {\mathcal M}^3}\frac{[(K-2-q^2)x+3q\bar Am]}{\left\{
-W^2+3y(y-m)+3i[x(q^2-1)+q\bar Am]\right\}}\ .
\end{eqnarray}

The limiting cases of Kerr, Schwarzschild, C metric and Rindler spacetimes are discussed below.

\subsection{Limiting cases}

\subsubsection{Kerr: $A=0$}

Consider the radial equation (\ref{radeqkerr}). One can find analytic solutions of this equation for $s=+2$ given by
\beq
R(r)=\frac{c_3r^3+c_2r^2+c_1r+c_0}{\Delta^2}e^{-i\omega \tilde r}, \qquad \tilde r= \int \frac{r^2+a^2-\frac{am}{\omega}}{\Delta} \, \rmd r\ .
\eeq
Substituting into Eq. (\ref{radeqkerr}) one obtains a 5$^{th}$ degree polynomial equation in $r$ which must be identically zero.
This system is compatible assuming the following constraint:
\beq
P_2(x)+P_4(y)+
C=0\ ,
\eeq
where $x={\mathcal M}\omega$, $y=a\omega$ and
\begin{eqnarray}
P_2(x)&=&9x^2 \nonumber \\
P_4(y)&=&9y^2(y-m)^2-2y(K-2)[(5K-7)y -(5K-13)m] \ .
\end{eqnarray}
We have 
\begin{eqnarray}
C=[(K-2)(K-3)]^2\equiv  W^2\ .
\end{eqnarray}

The solution is implicitly given by
\begin{eqnarray}
\frac{c_0}{{\mathcal M}^4}&=& \frac{c_1}{ {\mathcal M}^3} \frac{[i\bar a(y-m)-1]}{K-2}-\frac{c_2}{ {\mathcal M}^2} \frac{\bar a^2}{K-2}\nonumber \\
\frac{c_3}{{\mathcal M}}&=&-\frac{c_2}{{\mathcal M}^2} \frac{x}{K-2}
\end{eqnarray}
with $c_1$ arbitrary and $c_2$ given by
\begin{eqnarray}
\frac{c_2}{{\mathcal M}^2}&=&2i\frac{c_1}{ {\mathcal M}^3}\frac{(K-2)x}{[
-W^2+3y(y-m)-3ix]}\ .
\end{eqnarray}

\subsubsection{Schwarzschild: $A=0$, $a=0$}

Consider the radial equation (\ref{radeqschw}). One can find analytic solutions of this equation for $s=+2$ given by
\beq
R(r)=\frac{c_3r^3+c_2r^2+c_1r+c_0}{(r^2-2{\mathcal M}r)^2}e^{-i\omega \tilde r}, 
\eeq
with
\beq
\tilde r= \int \frac{r}{r-2{\mathcal M}} \, \rmd r=r+2{\mathcal M}\log (r-2{\mathcal M})\ .
\eeq
Substituting into Eq. (\ref{radeqschw}) one obtains a 4$^{th}$ degree polynomial equation in $r$ which must be identically zero.
This system is compatible assuming the following constraint:
\beq
P_2(x)+C=0\ ,
\eeq
where $x={\mathcal M}\omega$ and
\begin{eqnarray}
P_2(x)&=&9x^2\ .
\end{eqnarray}
We have 
\begin{eqnarray}
C=[(K-2)(K-3)]^2\equiv  W^2\ .
\end{eqnarray}

The solution is implicitly given by
\begin{eqnarray}
\frac{c_0}{{\mathcal M}^4}&=& -\frac{c_1}{ {\mathcal M}^3} \frac{1}{K-2}\nonumber \\
\frac{c_3}{{\mathcal M}}&=&-\frac{c_2}{{\mathcal M}^2} \frac{x}{K-2}
\end{eqnarray}
with $c_1$ arbitrary and $c_2$ given by
\begin{eqnarray}
\frac{c_2}{{\mathcal M}^2}&=&-2i\frac{c_1}{ {\mathcal M}^3}\frac{(K-2)x}{
W^2+3ix}\ .
\end{eqnarray}

\subsubsection{C metric: $a=0$}

Consider the radial equation (\ref{radeqcmet}). One can find analytic solutions of this equation for $s=+2$ given by
\beq
R(r)=\frac{c_3r^3+c_2r^2+c_1r+c_0}{[(1-A^2r^2)(r^2-2{\mathcal M}r)]^2}e^{-i\omega \tilde r}, \qquad \tilde r= \int \frac{r}{(1-A^2r^2)(r-2{\mathcal M})} \, \rmd r\ .
\eeq
Substituting into Eq. (\ref{radeqcmet}) one obtains a 6$^{th}$ degree polynomial equation in $r$ which must be identically zero.
This system is compatible assuming the following constraint:
\beq
P_2(x)+C=0\ ,
\eeq
where $x={\mathcal M}\omega$ and
\begin{eqnarray}
P_2(x)&=&9x^2 .
\end{eqnarray}
We have 
\begin{eqnarray}
C&=& [(K-2)(K-3)-3{\bar A}^2]^2\equiv  W^2 \ .
\end{eqnarray}

The solution is implicitly given by
\begin{eqnarray}
\frac{c_0}{{\mathcal M}^4}&=&- \frac{c_1}{ {\mathcal M}^3} \frac{1}{K-2}\nonumber \\
\frac{c_3}{{\mathcal M}}&=& \frac{c_1}{{\mathcal M}^3} \frac{\bar A^2}{K-2}+\frac{c_2}{{\mathcal M}^2} \frac{\bar A^2-x}{K-2}
\end{eqnarray}
with $c_1$ arbitrary and $c_2$ given by
\begin{eqnarray}
\frac{c_2}{{\mathcal M}^2}&=&-2i\frac{c_1}{ {\mathcal M}^3}\frac{(K-2)x}{W^2+3ix}\ .
\end{eqnarray}

\subsubsection{Rindler: ${\mathcal M}=0$, $a=0$}

Consider the radial equation (\ref{radeqrind}). One can look for analytic solutions of this equation for $s=+2$ given by
\beq
R(r)=\frac{c_3r^3+c_2r^2+c_1r+c_0}{(1-A^2r^2)^2}e^{-i\omega \tilde r}, 
\eeq
with
\beq
\tilde r= \int \frac{r^2}{1-A^2r^2} \, \rmd r=\frac{1}{A^2}\left[\,-r +\frac{1}{A}\log \sqrt{\left| \frac{1+Ar}{1-Ar}\,\, \right|} \right]\ .
\eeq
Substituting into Eq. (\ref{radeqrind}) one obtains a 5$^{th}$ degree polynomial equation in $r$ which must be identically zero.
This system admits only the trivial solution $c_3=c_2=c_1=c_0=0$ and therefore in this case there exists no algebraically special solutions.

\section{Concluding remarks}

A master equation for the gauge- and tetrad-invariant first-order massless perturbations of any spin $s\leq2$ on the spinning C metric background spacetime has been obtained and separated.
We have studied superradiance in this case and have shown that the situation is very similar to the Kerr spacetime; in particular, we have demonstrated that there is no coupling of spacetime acceleration and rotation in superradiant modes. 
This investigation offers the possibility of achieving a better understanding of perturbations of black hole spacetimes within this larger family; moreover, our work may be relevant to string theory, where the spinning C metric spacetime is of interest for other reasons.

\appendix

\section{Geometrical properties of the spinning C metric: an overview}

We list here the most important properties of the spinning C metric written in the form (\ref{SCmmetric}) following Podolsky and Griffiths \cite{podgriff1,podgriff2}.

\begin{itemize}

\item {\it Ring singularity at $r=0$}

The solution is characterized by the presence of a Kerr-like ring singularity at $r=0$, $\theta=\pi/2$, as shown by Podolsky and Griffiths \cite{podgriff1,podgriff2}. 

\item {\it Killing horizons}

Surfaces on which $Q=0$ are Killing horizons: $r=r_\pm$, $r=r_A\equiv 1/A$. The expressions for $r_\pm$ are identical to those for the locations of the outer and inner horizons of the non-accelerating Kerr black hole. The additional horizon at $r=r_A$, which is already familiar in the context of the C metric, is an acceleration horizon.

\item {\it Conformal infinity}

The vanishing of the conformal factor $\Omega$ corresponds to conformal infinity. 
If $\theta\in(0,\pi/2)$ the latter is  given by $r=1/(A\cos\theta)$, so that the range of allowed values of $r$ turns out to be $r\in(0,1/(A\cos\theta))$; if instead $\theta\in(\pi/2,\pi)$ conformal infinity is not reached even for $r\to\infty$, so that we may take $r\in(0,\infty)$.

\item {\it Conical singularities}

Conical singularities generally occur on the axis at both $\theta=0$ and $\theta=\pi$. However, by specifying the range of $\phi$ appropriately, the singularity on one half of the axis can be removed. 
For example, that on $\theta=\pi$ is removed by taking $\phi\in[0,2\pi/\sigma_+)$  corresponding to a pair of strings providing the necessary acceleration by connecting the sources to infinity. 
Alternatively, the singularity on $\theta=0$ can be removed by taking $\phi\in[0,2\pi/\sigma_-)$  corresponding to a strut between the sources. 

How a restriction in the allowed values of the $\phi$ coordinate can be used to remove the singularity can be easily seen
by considering the limiting form of the metric on the axis. For instance, on $\theta=0$ the latter reduces to
\beq
g_{\alpha\beta}\to \frac{1}{(1-Ar)^2}{\rm diag} \left[-\frac{Q}{(r^2+a^2)},\frac{(r^2+a^2)}{Q},\frac{(r^2+a^2)}{\sigma_- },0 \right]+O(\theta^2);
\eeq
more precisely $g_{\phi\phi}=(r^2+a^2)/(1-Ar)^2\sigma_-\theta^2+o(\theta^2)$; therefore the (limiting) metric induced on a $t=$const, $r=$const sphere  is conformal to 
\beq
{}^{(2)}\rmd s^2= \rmd \theta^2 + \theta^2 \sigma_-^2 \rmd \phi^2
\eeq
which coincides with that of a right cone once $\phi\in[0,2\pi/\sigma_-)$, as stated above. Similar considerations hold in the case $\theta=\pi$ with
$\sigma_-$ replaced by $\sigma_+$.

\item {\it Ergosurfaces}

The norm of the timelike Killing vector $\partial_t$ is given by 
\begin{equation}
\partial_t \cdot \partial_t = -\frac{Q-a^2P\sin^2\theta}{\Omega^2\Sigma}\ ,
\end{equation}
so that $\partial_t$ is timelike for $Q-a^2P\sin^2\theta>0$.
The vanishing of this quantity implicitly defines the ergoregions \cite{farzim1,farzim2}.

\item {\it Choice of the parameters and the region of validity of the coordinates}

We limit our attention to the spherical corona ${\mathcal C}$: 
$r_+<r<r_A$ in which $Q(r)>0$. This choice implies $1/A >r_+$, that is the condition
$A<A_{*}\equiv 1/r_+=r_-/a^2$. The latter restriction  also  implies $P(\theta)>0$ for all $\theta \in [0,\pi]$.
Finally either one of the two conical singularities is assumed to be removed and this corresponds to a limitation for the values  of $\phi$ as explained above. There is no need now to specify if the removed one corresponds to $\theta=0$ or $\theta=\pi$.

\end{itemize}

\section{Liouvillian solutions to the radial equation}

We apply here the well-known Kovacic algorithm \cite{kovacic} to the radial equation (\ref{radeq}) governing the perturbations in order to find the closed-form, i.e. Liouvillian, solutions to it.
Roughly speaking, the set of Liouvillian functions includes the usual elementary functions such as
exponential, trigonometric, logarithmic functions, etc., but not generic hypergeometric functions or other special functions.
An algorithm for computing solutions of second order linear ordinary differential equations in terms of special functions has been proposed by Bronstein and Lafaille \cite{bronstein} and can be in turn applied as well to the present case.

Let us examine the pole structure of the radial potential (\ref{radiale1}), rewritten in the form
\beq
V_{\rm (rad)}(r)=a_2r^2+a_1r+a_0+\frac{b_A^-}{r-r_A}+\frac{b_A^+}{r+r_A}+\frac{b_+}{r-r_+}+\frac{b_-}{r-r_-}\ ,
\eeq
where $a_2=-2A^2(1+s)(1+2s)$, $a_1=-{\mathcal M}a_2$, $a_0=2K-\omega^2/A^2-2is{\mathcal M}\omega$ and 
\begin{eqnarray}
b_A^\pm&=&\frac{\kappa(r_A)^2}{X_A^\pm}-is\kappa (r_A)\ , \qquad
b_\pm=\frac{\kappa(r_\pm)^2}{X_\pm}-is\kappa(r_\pm)\ ,
\end{eqnarray}
being 
\beq
X_A^\pm=\pm2\sigma_\pm r_A\ , \qquad
X_\pm=\frac{(-r_\pm^2+a^2)(A^2r_\pm^2-1)}{r_\pm}\ .
\eeq
$X_A^\pm$ and $X_\pm$ can be cast in a more symmetric form. In fact we find
\beq
X_A^\pm =\pm \frac{2}{r_A}(\pm r_A-r_+)(\pm r_A-r_-), \quad X_\pm=\mp \frac{2}{r_A} \frac{r_+ -r_-}{2r_A}(r_\pm -r_A)(r_\pm +r_A)\ . 
\eeq

Eq. (\ref{radeq}) can then be cast in the form 
\beq
\label{radeqNF}
\frac{\rmd^2y(r)}{\rmd r^2}=W_{\rm (rad)}(r)y(r)
\eeq
by the scaling
\beq
R=yQ^{-(1+s)/2}\ ,
\eeq
where the function $W_{\rm (rad)}(r)$ is given by
\begin{eqnarray}
\label{Wdef}
W_{\rm (rad)}(r)&=&\frac{c_A^-}{(r-r_A)^2}+\frac{c_A^+}{(r+r_A)^2}+\frac{c_+}{(r-r_+)^2}+\frac{c_-}{(r-r_-)^2}\nonumber\\
&&+\frac{d_A^-}{r-r_A}+\frac{d_A^+}{r+r_A}+\frac{d_+}{r-r_+}+\frac{d_-}{r-r_-}\ ,	
\end{eqnarray}
where
\begin{eqnarray}
c_A^\pm&=&-\frac{1-s^2}{4}-\frac{b_A^\pm}{X_A^\pm}\ , \nonumber\\
c_\pm&=&-\frac{1-s^2}{4}-\frac{b_\pm}{X_\pm}\ , \nonumber\\
d_A^\pm&=&\pm\frac{2i\omega sr_A}{X_A^\pm}\mp\frac{A(1+s)^2}{4}+\frac{2s(1+s)}{X_A^\pm}(1\pm A{\mathcal M})-\frac{2K}{X_A^\pm}+\frac{\omega^2}{A^2X_A^\pm}\nonumber\\
&&+\frac{2}{X_A^\pm{}^2}\left\{
\frac{\kappa(r_+)^2}{X_+}(1\pm Ar_-)+\frac{\kappa(r_-)^2}{X_-}(1\pm Ar_+)\right.\nonumber\\
&&\left.-\frac{\kappa(r_A)^2}{X_A^\pm}\left[2\pm\frac{AX_A^\pm}{2}\pm2A{\mathcal M}\left(1-\frac{X_A^\pm}{X_A^\mp}\right)\right]
\right\}\ , \nonumber\\
d_\pm&=&-\frac{2i\omega sr_\pm}{X_\pm}\pm\frac{(1+s)^2}{2(r_+-r_-)}\pm\frac{s(1+s)A^2r_\pm }{X_\pm}(r_+-r_-)-\frac{2K}{X_\pm}-\frac{\omega^2}{A^2X_\pm}\nonumber\\
&&\pm\frac{1}{X_\pm{}^2}\left\{
(A^2r_\pm^2-1)\left[\frac{\kappa(r_-)^2}{X_-}-\frac{\kappa(r_+)^2}{X_+}\right]\right.\nonumber\\
&&\left.+A(r_+-r_-)\left[-2Ar_\pm\frac{\kappa(r_\pm)^2}{X_\pm}+\left(\frac{1+Ar_\pm}{X_A^-}-\frac{1-Ar_\pm}{X_A^+}\right)\kappa(r_A)^2\right]
\right\}\ . 
\end{eqnarray}

The Kovacic algorithm applies to a general second order ordinary differential equation (DE) of the form (\ref{radeqNF}) when 
$W_{\rm (rad)}(r)$  is a given element of $\mathbb{C}(r)$, the field of rational functions with coefficients in the field of complex numbers.
Kovacic proved that all the Liouvillian solutions to Eq. (\ref{radeqNF}) are three mutually exclusive types.
Furthermore, for each type, he provided an algorithm which decides if a Liouvillian solution exists
for that type and constructs the solution if it does exist. The types are as follows:

\begin{itemize}

\item[Type 1.]
The DE has a solution of the form $e^{\int\tilde\Omega\rmd r}$, where $\tilde\Omega\in\mathbb{C}(r)$.

\item[Type 2.]
The DE has a solution of the form $e^{\int\tilde\Omega\rmd r}$, where $\tilde\Omega$ is algebraic over $\mathbb{C}(r)$ of degree 2 (i.e., $\tilde\Omega$ is a solution of a polynomial equation of degree 2 with coefficients in $\mathbb{C}(r)$), and type 1 does not hold.

\item[Type 3.]
All solutions of the DE are algebraic over $\mathbb{C}(r)$ of degree 2, and types 1 and 2 do not hold.

\end{itemize}

If none of the previous cases applies, the DE has no Liouvillian solution.

The construction of each type of Liouvillian solution is derived from the pole structure of the rational function.
Kovacic also discussed some conditions that are necessary for cases 1, 2 or 3 to hold. 
For type 1 it is necessary that every pole of $W_{\rm (rad)}(r)$ have even order or else order 1, and that the pole at $\infty$ have either even order or order greater than 2.
For type 2 a necessary condition is that $W_{\rm (rad)}(r)$ must have at least one pole that has either odd order greater than 2 or order 2.
Finally, for type 3 the order of a pole of $W_{\rm (rad)}(r)$ cannot exceed 2, and the order of the pole at $\infty$ must be at least 2.

Let us discuss in detail the case 1. The remaining cases 2 and 3 apply as well and can be treated similarly.
Note that type 3 is instead empty for all black hole perturbations considered by Couch and Holder \cite{couch} including the Kerr case, since the necessary condition for this type is not fulfilled.

For a fixed rational function $W_{\rm (rad)}(r)$, type 1 Liouvillian solutions of Eq. (\ref{radeqNF}) are determined in the following way. A finite set of constants $d$ is derived algorithmically from the pole structure of $W_{\rm (rad)}(r)$
and for each $d$ a function $\tilde\Omega(r)$ is also constructed. If any $d$ is a
non-negative integer $n$ and, for the corresponding $\tilde\Omega(r)$, the differential equation
\beq
\label{eqPdir}
\frac{\rmd^2{\mathcal P}(r)}{\rmd r^2}+2\tilde\Omega\frac{\rmd {\mathcal P}(r)}{\rmd r} + \left[\frac{\rmd\tilde\Omega}{\rmd r}+\tilde\Omega^2-W_{\rm (rad)}(r)\right]{\mathcal P}(r)=0\ 
\eeq
has a polynomial solution ${\mathcal P}$ of degree $n$ then Eq. (\ref{radeqNF}) has a Liouvillian solution given by
\beq
\label{ysol}
y={\mathcal P}e^{\int\tilde\Omega\rmd r}\ .
\eeq
If this is not the case for any $d$ then Eq. (\ref{radeqNF}) has no type 1 solutions.

Since $W_{\rm (rad)}(r)$ depends on different independent parameters (black hole parameters, spacetime acceleration, separation constants), several different pole structures may occur by imposing conditions on the parameters which alter the order of poles. For each different pole structure, in general, $d$ and $\tilde\Omega$ are expressed in terms of the parameters. Solutions to Eq. (\ref{radeqNF}) may be obtained by requiring $d=n$, which then is just a constraint on the parameters, and finding all sets of parameters for which Eq. (\ref{eqPdir}) has a polynomial solution of degree $n$.

For all possible pole structures of (\ref{Wdef}), all functions $\tilde\Omega$ generated by the algorithm for type 1
have the form
\beq
\tilde\Omega=\frac{\alpha_+}{r-r_+}+\frac{\alpha_-}{r-r_-}+\frac{\alpha_{A}^+}{r+r_A}+\frac{\alpha_{A}^-}{r-r_A}\ ,
\eeq
where
\beq
\alpha_{\pm}=\frac12\left[1+\delta_\pm\sqrt{1+4c_\pm}\right]\ , \qquad
\alpha_A^{\pm}=\frac12\left[1+\delta_A^\pm\sqrt{1+4c_A^\pm}\right]\ ,
\eeq
the quantities $\delta_+$, $\delta_-$, $\delta_A^+$ and $\delta_A^-$ taking independently values $\pm1$.

The necessary conditions for type 1 solutions are fulfilled if the following constraint is imposed
\beq
\label{cond1}
d_++d_-+d_A^++d_A^-=0\ ,
\eeq
implying that the order of the pole at infinity has order 2.

The algorithmic condition $d=n$ is given by 
\beq
\label{cond2}
\alpha_\infty-\alpha_+-\alpha_--\alpha_{A}^+-\alpha_{A}^-=n\ ,
\eeq
where
\beq
\alpha_{\infty}=\frac12\left[1+\delta_\infty\sqrt{1+4(c_++c_-+c_A^++c_A^-)}\right]\ , \qquad \delta_\infty=\pm1\ .
\eeq

For every pole structure, we construct the Liouvillian solutions to Eq. (\ref{radeqNF}) by deriving a consistent
recursion relation which generates the finite set of coefficients in the polynomial ${\mathcal P}$. 
Following Couch and Holder \cite{couch}, to obtain a recursion relation with the fewest number of terms we take, without loss of generality, ${\mathcal P}$ to be expanded about a pole of nonzero order, say $r_+$, as follows
\beq
{\mathcal P}=\sum_{k=0}^n a_k(r-r_+)^k\ , \qquad a_n\not=0\ .
\eeq
After substituting this expression for ${\mathcal P}$ into Eq. (\ref{eqPdir}) we obtain in the general case, when all three poles are present, the following  recursion relation
\begin{eqnarray}
\label{recureq}
0&=&[H_0+(k-2)(k-3-2n+2\alpha_\infty)]a_{k-2}\nonumber\\
&&+\{H_1+G_0[H_0+(k-1)(k-2-2n+2\alpha_\infty)]+2L_0(k-1)\}a_{k-1}\nonumber\\
&&+[g_+K_0+k(k-1-2n+2\alpha_\infty)J_0+2k(G_0L_0+L_1)]a_{k}\nonumber\\
&&+(k+1)K_0(k+2\alpha_+)a_{k+1}\ , \qquad k=0,\ldots,n+1\ ,
\end{eqnarray}
being understood that $a_i=0$ if $i\leq-1$ or $i\geq n+1$, plus the further constraint on the parameters
\beq
\label{cond3}
0=(r_+-r_-)d_-+(r_++r_A)d_A^++(r_+-r_A)d_A^-\ .
\eeq
The constants entering Eq. (\ref{recureq}) are listed below
\begin{eqnarray}
G_0&=&3r_+-r_-\ , \nonumber\\
H_0&=&-(r_+-r_-)g_--(r_++r_A)g_A^+-(r_+-r_A)g_A^-\ , \nonumber\\
H_1&=&(r_+-r_-)^2g_-+(r_++r_A)^2g_A^++(r_+-r_A)^2g_A^-\ , \nonumber\\
K_0&=&(r_+-r_-)(r_+^2-r_A^2)\ , \nonumber\\
J_0&=&3r_+^2-2r_+r_--r_A^2\ , \nonumber\\
L_0&=&-(r_+-r_-)\alpha_--(r_++r_A)\alpha_A^+-(r_+-r_A)\alpha_A^-\ , \nonumber\\
L_1&=&(r_+-r_-)^2\alpha_-+(r_++r_A)^2\alpha_A^++(r_+-r_A)^2\alpha_A^-\ ,
\end{eqnarray}
where 
\begin{eqnarray}
g_\pm&=&-d_\pm+2\alpha_\pm\left[\pm\frac{\alpha_\mp}{r_+-r_-}+\frac{\alpha_A^+}{r_\pm+r_A}+\frac{\alpha_A^-}{r_\pm-r_A}\right]\ , \nonumber\\
g_A^\pm&=&-d_A^\pm-\alpha_A^\pm\left[2\frac{\alpha_+}{r_+\pm r_A}+2\frac{\alpha_-}{r_-\pm r_A}\pm\frac{\alpha_A^\mp}{r_A}\right]\ .
\end{eqnarray}

Liouvillian solutions (\ref{ysol}) to Eq. (\ref{radeqNF}) are thus given by
\beq
y={\mathcal P}(r-r_+)^{\alpha_+}(r-r_-)^{\alpha_-}(r+r_A^+)^{\alpha_+}(r_A-r)^{\alpha_A^-}\ ,
\eeq
implying that 
\beq
R={\mathcal P}r_A^{1+s}(r-r_+)^{\alpha_+-\frac{1+s}2}(r-r_-)^{\alpha_--\frac{1+s}2}(r+r_A^+)^{\alpha_A^+-\frac{1+s}2}(r_A-r)^{\alpha_A^--\frac{1+s}2}\ .
\eeq

A detailed discussion of such solutions which takes into account the constraints (\ref{cond1}), (\ref{cond2}) and (\ref{cond3}) by specializing the parameters involved is beyond the scope of the present paper.

\section{Null geodesics}

For completeness we include here also the separated equations for null geodesics.
As in the case of the non spinning C metric only null geodesics can be separated because of the existence of a conformal Killing tensor.
The latter has a simple form when expressed in the NP frame (\ref{frame}):
\beq
P_{\alpha\beta}=\frac{\Sigma}{4\Omega^2}[l_{(\alpha}n_{\beta)}+m_{(\alpha}{\bar m}_{\beta)}]\ .
\eeq
The result is the following
\begin{eqnarray}
\frac{\rmd t}{\rmd \lambda} &=& \frac{\Omega^2}{\Sigma}\left\{ \frac{(r^2+a^2)[E(r^2+a^2)-La]}{Q}+a\frac{L-aE\sin^2\theta}{P}\right\},\nonumber \\
\frac{\rmd r}{\rmd \lambda} &=& \pm \frac{\Omega^2}{\Sigma}\left\{[E(r^2+a^2)-aL]^2-C Q\right\}^{1/2},\nonumber \\
\frac{\rmd \theta}{\rmd \lambda} &=& \pm \frac{\Omega^2}{\Sigma}\left[CP-\frac{(L-aE\sin^2\theta)^2}{\sin^2\theta} \right]^{1/2},\nonumber \\
\frac{\rmd \phi}{\rmd \lambda} &=& \frac{\Omega^2}{\Sigma}\left\{ \frac{a[E(r^2+a^2)-La]}{Q}+\frac{L-aE\sin^2\theta}{P\sin^2\theta} \right\},
\end{eqnarray}
where $C$ is a separation constant related to the conformal Killing tensor by 
\beq
P_{\alpha\beta}\frac{\rmd x^\alpha}{\rmd \lambda}\frac{\rmd x^\beta}{\rmd \lambda}=\frac{C}{2}\ .
\eeq

Let us consider as an example the case of circular geodesics on the equatorial plane, i.e. orbits at $r=r_0=$fixed and $\theta=\pi/2=$fixed. This means
\beq
C=(L-aE)^2\ ,
\eeq
from the $\theta-$equation and
\beq
\frac{[E(r_0^2+a^2)-aL]}{\sqrt{Q_0}}=\pm (L-aE)=\pm \sqrt{C}\ ,
\eeq
from the $r-$equation. Moreover inserting these conditions in the original (second order) geodesic equations one finds the allowed values of $r_0$ as the solutions of the following $5^{th}$ order equation in $\rho=r/{\mathcal M}$:
\begin{eqnarray}
\label{eqcirc}
\rho^{5}{\bar A}^{4}+ 2{\bar A}^{2}\left(2 -\sigma_0\right) 
\rho^{4}+ \left[ \sigma_0^2-6{\bar A}^{2}\right] \rho^{3} + 2\left( \sigma_0-4 \right)\rho^{2}+
9\rho-4{\bar a}^{2}=0\ .
\end{eqnarray}
When $\bar A=0$, i.e. in the case of Kerr spacetime, Eq. (\ref{eqcirc}) reduces to 
\begin{eqnarray}
\rho (\rho-3)^2-4{\bar a}^{2}=0\ ,
\end{eqnarray}
with solutions \cite{chandra}
\beq
\rho_{_{(K\pm)}}=4\cos^2\beta_\pm\ , \qquad \beta_\pm=\frac13\arccos(\pm {\bar a})\ .
\eeq
To second order in $\bar A$ the solutions of Eq. (\ref{eqcirc}) are given by
\beq
\rho_\pm=\rho_{_{(K\pm)}}\left[1-\frac16(\rho_{_{(K\pm)}}-1)\rho_{_{(K\pm)}}^2\,{\bar A}^2\right]+O({\bar A}^4)\ .
\eeq


\begin{thebibliography}{00}

\bibitem{ES}
H. Stephani, D. Kramer, M. A. H. MacCallum, C. Hoenselaers, and E. Herlt,  
\textit{Exact Solutions of Einstein's Field Equations}, 2nd ed.
(Cambridge Univ.\ Press, Cambridge, 2003).

\bibitem{podgriff1}
J. Podolsk\'y and J. B. Griffiths,
Class.\ Quantum\ Grav. {\bf 22}, 3467 (2005).

\bibitem{podgriff2}
J. Podolsk\'y and J. B. Griffiths,
Phys.\ Rev.\ D {\bf 73}, 044018 (2006).

\bibitem{Teukolsky}  
S. A. Teukolsky,  
Astrophys.\ J. {\bf 185}, 635 (1973).

\bibitem{barpre}
J. M. Bardeen and W. H. Press,  
J.\ Math.\ Phys. {\bf 14}, 7 (1973).

\bibitem{NP}  
E. T. Newman and R. Penrose, 
J.\ Math.\ Phys. {\bf 3}, 566 (1962).

\bibitem{chandra}  
S. Chandrasekhar,
{\it The Mathematical Theory of Black Holes\/}
(Oxford Univ.\ Press, New York, 1983). 

\bibitem{Prestidge}
T. Prestidge, 
Phys.\ Rev. {\bf 58}, 124022 (1998).

\bibitem{kinnersley}
W. Kinnersley,  
J. Math.\ Phys. {\bf 10}, 1195 (1969).

\bibitem{guven}
R. G\"uven, 
Phys.\ Rev. D {\bf 22}, 2327 (1980).

\bibitem{GHP} 
R. Geroch, A. Held, and R. Penrose, 
J.\ Math.\ Phys. {\bf 14}, 874 (1973).

\bibitem{bcjr}
D. Bini, C. Cherubini, R. T. Jantzen, and R. Ruffini,
Prog.\ Theor.\ Phys. {\bf 107}, 1 (2002).

\bibitem{TeukTN}
D. Bini, C. Cherubini, and R. T. Jantzen,
Class. Quantum Grav. {\bf 19}, 1 (2002).

\bibitem{JAC}
J. D. Bekenstein  and  M. Schiffer, 
Phys.\ Rev.\ D {\bf 58}, 064014 (1998).

\bibitem{zeldovich}
Ya. B. Zel'dovich, 
Zh. Eksp. Teor. Fiz. {\bf 62}, 2076 (1971) [Sov. Phys. JETP {\bf 35}, 1085 (1971)].

\bibitem{chersucci1}
C. Cherubini, F. Federici, S. Succi, and M. P. Tosi,
Phys.\ Rev.\ D {\bf 72}, 084016 (2005).

\bibitem{FC} 
E. D. Fackerell and R. G. Crossman,
J.\ Math.\ Phys. {\bf 18}, 1849 (1977).

\bibitem{novello}
B. D. B. Figueiredo and M. Novello, 
J.\ Math.\ Phys. {\bf 34}, 3121 (1993).

\bibitem{farzim1}
H. Farhoosh and R. L. Zimmerman, 
Phys.\ Rev.\ D {\bf 22}, 797 (1980).

\bibitem{farzim2}
H. Farhoosh and R. L. Zimmerman, 
Phys.\ Rev.\ D {\bf 21}, 2064 (1980).

\bibitem{kovacic}
J. J. Kovacic, 
J. Symb. Comput. {\bf 2}, 3 (1986).

\bibitem{bronstein}
M. Bronstein and S. Lafaille,
in {\it Proceedings of ISSAC 2002}, ACM Press, pp 23-28.

\bibitem{couch}
W. E. Couch and C. L. Holder,
J. Math. Phys. {\bf 22}, 1457 (1981).

\end{thebibliography}
\end{document}